

\input harvmac.tex
\noblackbox

\def\pmb#1{\setbox0=\hbox{#1}%
  \kern-.025em\copy0\kern-\wd0
  \kern.05em\copy0\kern-\wd0
  \kern-.025em\raise.0433em\box0 }

\def\zhat{{\bf \hat{z}}}
\def\chat{{\bf \hat{c}}}
\def\rvec{{\bf r}}
\def\rp{{\bf r}_{\perp}}

\def\qhat{{\bf \hat{q}}_{\perp}}
\def\perphat{\zhat\times\qhat}

\def\qp{q_{\perp}}
\def\qpvec{{\bf q}_{\perp}}

\def\lab{\lambda_{ab}}

\def\beff{B_3\xi_{\parallel}}
\def\xip{\xi_{\parallel}}


\Title{}{\vbox{\centerline{Translational Correlations in the Vortex
Array}
     \vskip2pt\centerline{at the Surface of a Type-II Superconductor}}}


{\baselineskip= 16pt plus 2pt minus 1pt\centerline{M. CRISTINA MARCHETTI}
\footnote{}
\smallskip\centerline{Physics Department}
\centerline{Syracuse University}
\centerline{Syracuse, NY 13244}
\medskip
\centerline{DAVID R. NELSON}
\smallskip\centerline{Lyman Laboratory of Physics}
\centerline{Harvard University}
\centerline{Cambridge, MA 01238}}

\vskip .1in

We discuss the statistical mechanics of magnetic flux lines in a
finite-thickness slab of type-II superconductor. The long
wavelength properties of a flux-line liquid in a slab geometry
are described by a hydrodynamic free energy that incorporates the
boundary conditions on the flux lines at the sample's surface
as a surface contribution to the free energy. Bulk and surface
weak disorder are modeled via Gaussian impurity potentials.
This free energy is used to evaluate the two-dimensional
structure factor of the flux-line tips at the sample surface.
We find that surface
interaction always dominates in determining the
decay of translational correlations in the asymptotic long-wavelength limit.
On the other hand, such large length scales have not been
probed by the decoration experiments. Our results indicate
that the translational correlations
extracted from the analysis of the Bitter patterns are indeed
representative of behavior of flux lines in the bulk.

\Date{12/7/92}

\newsec{Introduction}

The nature of the ordering of the magnetic flux array in the mixed
state of high-temperature copper-oxide superconductors has received
considerable experimental and theoretical attention in the last
few years.
It has been shown that fluctuations are important in these
materials and can lead to a number of new phases or regimes of
the flux array, including entangled flux
liquids, hexatic flux liquids
\nref\mcmrev{M.C. Marchetti and D.R. Nelson,
Physica C {\bf 174}, 40 (1991).}\nref\drnrev{For a review see:
D.R. Nelson, in {\it Phenomenology
and Applications of High-Temperature Superconductors}, K.S. Bedell et
al. eds. (Addison-Wesley, 1992).}\refs{\mcmrev ,\drnrev}
and hexatic vortex glasses
\nref\ffhuse{D.S. Fisher, M.P.A. Fisher, and D. Huse, Phys. Rev.
B{\bf 43}, 130 (1991).}
\nref\dsfrev{D.S. Fisher, in {\it Phenomenology
and Applications of High-Temperature Superconductors}, K.S. Bedell et
al. eds. (Addison-Wesley, 1992).}\refs{\ffhuse ,\dsfrev}.
Most experiments probe the properties of the flux array
indirectly by measuring bulk properties of the superconductors, such
as
transport, magnetization or mechanical dissipation.
At present direct measurements of the microscopic order of the magnetic
flux array are mainly limited to decoration experiments at low fields
\nref\deca{P.L. Gammel, D.J. Bishop, G.J. Dolan, J.R. Kwo,
C.A. Murray, L.F. Schneemeyer, and J.V. Waszczak, Phys. Rev. Lett.
{\bf 59}, 2592 (1987).}\nref\dolan{G.J. Dolan, G.V. Chandrasekar,
T.R. Dinger, C. Field, and F. Holtzberg, Phys. Rev. Lett. {\bf 62},
827 (1989).}\nref\dechex{C.A. Murray, P.L. Gammel,
D.J. Bishop, D.B. Mitzi, and A. Kapitulnik, Phys. Rev. Lett. {\bf 64},
2312 (1990).}\nref\grier{D.G. Grier, C.A. Murray, C.A. Bolle,
P.L. Gammel,
D.J. Bishop, D.B. Mitzi, and A. Kapitulnik, Phys. Rev. Lett.
{\bf 66}, 2270 (1991).}
\nref\gammel{P.L. Gammel, in {\it Phenomenology
and Applications of High-Temperature Superconductors}, K.S. Bedell et
al. eds. (Addison-Wesley, 1992), and references therein.}
\refs{\deca - \gammel}.
These experiments aim to extract information on the
vortex line configurations
in the bulk of the material by imaging the pattern of the magnetic
flux lines as they
emerge at the surface of the sample. The surface patterns are
determined by the interplay of thermal fluctuations
and impurity disorder. Both these mechanisms can be responsible for
disrupting translational and orientational order of vortex arrays
\nref\mcmdis{M.C. Marchetti and D. R. Nelson, Phys. Rev. B{\bf 41},
1910 (1990).}\nref\chud{E.M. Chudnovsky, Phys. Rev. B{\bf 43},
7831 (1991).}\refs{\mcmdis ,\chud}.
Surface roughness can also play a role in determining
the surface magnetic patterns. It is clear that to interpret
the experiments and assess whether one can indeed consider the
surface patterns as
representative of vortex line configurations in the bulk of the sample,
one needs to understand what are the relative effects
of bulk versus surface interactions and disorder in determining
the configuration of the vortex tips as they emerge at the surface.
Almost thirty years ago Pearl \nref\pearl{J. Pearl, J. App. Phys.
{\bf 37}, 4139 (1966).}\refs{\pearl} showed that the interaction
between the tips of straight flux lines at a superconductor-vacuum interface
decays as $1/r_{\perp}$ at large distances, with $r_{\perp}$
the distance between flux tips along the interface. In contrast,
the interaction between flux-line elements in bulk decays exponentially
at large distances.
For this reason Huse has questioned
the assumption that surface patterns are representative
of flux lines configurations in the bulk
and has argued that at low fields, where the intervortex
separation is large compared to the penetration length,
surface effects may play the dominant role
in determining the magnetic flux patterns seen at the surface
\nref\huse{D. A. Huse, Phys. Rev. B{\bf 46}, 8621 (1992).}
\refs{\huse}.

By analyzing flux decoration images one can extract quantitative
information on the decay of both translational and orientational
correlations of flux line tips at the sample surface.
In this paper we focus on the long-wavelength behavior of
translational correlations in the flux-line liquid phase.
This case may be relevant to the interpretation of
decoration experiments such as those by the
AT\&T group \refs{\gammel} for the following reason.
The decoration experiments are carried out by quenching the sample
in a small field from high to low temperature.
The observed flux patterns do not represent the equilibrium
configurations of
vortices at the low temperature where the decoration takes place,
but equilibrium configurations corresponding to a higher temperature,
$T_f$, where the flux array falls out of equilibrium.
While the value of $T_f$ is not known, there are indication
that it may be near the irreversibility line, $T_{irr}$,
which in turn has been found to be very close to $T_c$ at low fields
in the BSCCO samples used for the decorations \refs{\gammel}.
The experiments may then be probing a rather narrow range of
temperatures between $T_{irr}$ and $T_f$ where the flux array
is in the polymer-like state proposed by Nelson \nref\seung{D.R. Nelson
and S. Seung, Phys. Rev. B{\bf 39}, 9153 (1989).}\refs{\seung}.
The long wavelength static properties of such a glassy polymer
can be described in terms of a hydrodynamic flux-line
liquid free energy \refs{\mcmrev}.

Translational correlations of flux lines in three dimensions
are described by the density-density correlation function.
For convenience we consider the Fourier transform
of this correlation function in the plane normal
to the applied field,
\eqn\structz{n_0S(\qp,z_1,z_2)=
   \overline{<\delta n(\qpvec,z_1)~\delta n(-\qpvec,z_2)>}
  - \overline{<\delta n(\qpvec,z_1)><\delta n(-\qpvec,z_2)>},}
where $n(\qpvec,z)$ is the in-plane Fourier transform of the
coarse-grained flux-line density,
\eqn\density{n(\rp,z)=\sum_{i=1}^N~\delta(\rp-\rvec_i(z)).}
Here $\rvec_i(z)$ is the position of the $i$-th vortex in the $(x,y)$
plane as it wanders along the $\zhat$ ($\zhat\parallel{\bf H}$) axis
and $\delta n(\rp,z)=n(\rp,z)-n_0$
denotes the fluctuation of the local density
from its equilibrium value,
$n_0=B/\phi_0$, with $B$ the average induction at equilibrium
and $\phi_0$ the flux quantum.
The angular brackets denote a thermal average and the overbar
the average over quenched impurity disorder. The subtracted term
on the right hand side of Eq. \structz\ vanishes in the absence
of quenched disorder.

The two-dimensional structure function of a constant-$z$
cross-section of flux liquid is obtained from \structz\ by letting
$z_1=z_2$,
\eqn\structztwo{n_0S_2(\qp,z_1)=
   \overline{<|\delta n(\qpvec,z_1)|^2>}
  - |\overline{<\delta n(\qpvec,z_1)>}|^2.}
A factor of $n_0$ has been extracted from the definition of the
the structure function so that $S_2(\qp,z_1)\rightarrow 1$
as $\qp\rightarrow\infty$, for all values of $z_1$.

The behavior of the flux-line structure function \structz\
for bulk flux-line liquids in the
presence of both thermal fluctuations and quenched disorder
has been discussed before \nref\doussal{D. R. Nelson and
P. LeDoussal, Phys. Rev. B{\bf 42}, 10113 (1990).}\refs{\doussal}.
In an infinitely-thick superconductor the correlation of density
fluctuations at different heights only depends on the distance,
$|z_1-z_2|$.
The results are then most conveniently discussed
in terms of the full three-dimensional structure function
obtained by Fourier transforming \structz\ with respect to $z_1-z_2$,
\eqn\threestr{n_0S(\qp,q_z)=\overline{<|\delta n(\qpvec,q_z)|^2>}
      -|~\overline{<\delta n(\qpvec,q_z)>}~|^2.}
For simplicity we only discuss in this section the contribution to
the structure factor from thermal fluctuations, $S_T(\qp,q_z)$.
This is easily calculated in the long wavelength limit from
a simple hydrodynamic theory \refs{\mcmrev ,\doussal}, with the result,
\eqn\structw{S_T(\qp,q_z)={n_0k_BT\qp^2\over
   \qp^2c_L(\qp,q_z)+q_z^2c_{44}(\qp,q_z)},}
where $c_L(\qp,q_z)$ and $c_{44}(\qp,q_z)$ are the wave vector-dependent
compressional and tilt moduli of a bulk flux-line liquid, as given
for instance in Ref. \refs{\mcmrev}.
In the long wavelength limit one can invert the $q_z$ transform
by approximating the elastic constants in \structw\ with their
values at $q_z=0$, with the result,
\eqn\stzb{S_T(\qp,z_1-z_2)={n_0k_BT\over B_3(\qp)\xi_{\parallel}(\qp)}
         e^{-|z_1-z_2|/\xi_{\parallel}(\qp)},}
where
\eqn\xipar{\xi_{\parallel}(\qp)=\sqrt{{K(\qp)\over B_3(\qp)}}{1\over\qp}}
is the correlation length describing the decay of correlations in the
$z$ direction, with $B_3(\qp)=c_L(\qp,0)$ and
$K(\qp)=c_{44}(\qp,0)$.
The asymptotic long-wavelength ($\qp\rightarrow 0$)
behavior of the correlation function is
determined by the constant values of the moduli at
zero wavevector, $B_3(0)={B^2\over 4\pi}$ and
$K(0)={B^2\over 4\pi}+\tilde{c}_{44}(0)$,
where $\tilde{c}_{44}(0)$ is the tilt coefficient of a single
line at $q_z=0$. \nref\footd{As pointed out in Ref. \refs{\dsfrev},
the single-line tilt coefficient contains a term that
survives even in the limit of very large anisotropy and
was missed in much previous literature. Using the
results of \refs{\dsfrev}, one obtains
$\tilde{c}_{44}(0)=(n_0\phi_0^2/ 16\pi^2\lab^2)
\big[\ln(\gamma\kappa)/\gamma^2-(1/(2\gamma^2))+(1/ 2)\big]$.
Here $\lab$ and $\lambda_c=\gamma\lab$ are the penetration lengths
in the $ab$ plane and along the $\chat$ axis, respectively, with
$\gamma$ the anisotropy parameter. Also, $\kappa=\lab/\xi_{ab}$,
with $\xi_{ab}$ the coherence length in the $ab$ plane.}
\refs{\footd}.
On the other hand, the nonlocality of the
elastic constants is often important in the high-$T_c$
superconductors
even at small wave vectors because of the large
values of the penetration lengths \ref\foota{The reader may question
the consistency of approximating the elastic constants by their
values at $q_z=0$, while retaining their dependence on $\qp$.
It turns out that the main effect of the nonlocality
on the properties of interest here is that of reducing
the value of the bulk modulus at $\qp\sim 1/a_0$,
as compared to its value
at $\qp=0$. This effect is retained in the approximation used
here.}.

Correlations in any constant-$z$ cross section of the
bulk are described by the two-dimensional structure
factor defined in Eq. \structztwo\ and do not depend on $z_1$.
The contribution to this two-dimensional
structure factor from thermal fluctuations is immediately obtained
by letting $z_1=z_2$ in Eq. \stzb\ ,
\eqn\stb{\eqalign{S_{2T}(\qp)&=
{n_0k_BT \over B_3\xi_{\parallel}(\qp)}\cr
&={n_0k_BT \over\sqrt{KB_3}}~\qp.}}
According to a well known sum rule that relates the value
of the structure factor at zero-wavevector to the bulk modulus
of the liquid,
the denominator on the right hand side of the first equality of Eq. \stb\
can be interpreted as an effective two-dimensional bulk modulus.
This effective bulk modulus diverges in the
long wavelength limit, due to the
divergence of the correlation length $\xi_{\parallel}$.
As a consequence the two-dimensional
structure factor of any constant-$z$ cross-section of a bulk
flux-line liquid
vanishes linearly as $\qp\rightarrow 0$. As discussed elsewhere
\nref\drnjsp{D.R. Nelson, Physica A{\bf 177}, 220 (1991).}\refs{\doussal
,\drnjsp},
this behavior arises here from the
constraint that flux lines cannot start nor terminate inside the sample.

Decoration experiments measure correlations of flux lines in
a constant-$z$ cross-section at the surface of a sample of
finite thickness $L$.
Most researchers have implicitly assumed that these surface
correlations are representative of correlations in a constant-$z$
cross-section of bulk. If this is the case
Eq. \stb\ should describes the correlations of flux-line tips
at the surface as extracted from the decoration experiments.
On the other hand, it is clear that whenever
$\xi_{\parallel}\geq L$ finite-size effects may become
important. Since $\xi_{\parallel}(\qp)$ diverges as $\qp\rightarrow 0$
finite-size effects may in fact dominate the long wavelength behavior.
In addition we have mentioned that
in a finite size sample the boundary conditions
modify the pair interaction between flux lines at
the surface.
The boundary conditions are determined by the requirement
that on length scales larger than the penetration
length the spatially uniform magnetic field outside
the sample must equal the field of the vortex tips at the surface.
In other words the superconductor surface can be thought of as
the boundary between a strongly anisotropic magnetic medium where
the field is concentrated in the flux lines and an isotropic magnetic
medium (the vacuum).
At the surface the flux tips then behave like
magnetic monopoles of "charge" $\phi_0/2\pi$ and therefore
interact at large distances via a repulsive Coulomb-like
potential,
$V_s(r_{\perp})=(\phi_0/2\pi)^2(1/r_{\perp})$.
The two-dimensional correlations
of flux-line tips as they leave the sample
could then be very different from those in a constant-$z$ cross section
deep in the bulk. If in fact
the long-range surface interaction would dominate
in determining long-wavelength surface properties \refs{\huse},
then decorations would image the
configurations of a
two-dimensional liquid
of vortices interacting via a $1/r_{\perp}$ potential.
The long-wavelength bulk modulus of such a two-dimensional liquid
is $B_2(\qp)\simeq B^2/(4\pi\qp)$.
The corresponding two-dimensional structure factor
is given by $S_{2T}(\qp)=n_0k_BT/B_2(\qp)$ and again vanishes linearly
as $\qp\rightarrow 0$, but with a different slope from the result
of Eq. \stb .
A more detailed analysis is clearly needed to
discriminate between these two possibilities.

In this paper we describe a model that
can be used to study the long wavelength properties
of flux arrays in finite-size samples
incorporating the appropriate boundary conditions
for the flux lines at the sample surface.
We consider a slab of a uniaxial anisotropic type-II
superconductor. The field is applied along the $\chat$ axis of
the superconductor which is
chosen as the $z$ direction. The slab is infinite in the
$xy$ plane and has thickness $L$ in the $z$ direction.
The starting point for the calculation presented here is a model
hydrodynamic free energy that is a simplified version of
the full hydrodynamic free energy obtained in
\nref\mcm{M.C. Marchetti,
Physica C {\bf 200}, 155 (1992).}\refs{\mcm}.
In essence we neglect any nonlocality and spatial inhomogeneity
in the $z$ direction other than the presence of the sample
boundaries. The free energy is then written as the
sum of a surface free energy, that
incorporates the boundary conditions,
and the usual hydrodynamic free energy of a bulk flux liquid.
This free energy can be used as the starting point to study both
how the coupling to the bulk affects the surface properties
and how the bulk behavior is modified by the presence
of the boundaries.
In this paper we focus on translational correlations at the surface.
Our main result is the two dimensional structure factor at
one of the surfaces of the sample. We find that the contribution to this
structure factor from thermal fluctuations is given by,
\eqn\stans{S_{2T}(\qp,L)={n_0k_BT\over B_2^{eff}(\qp,L)},}
where $B_2^{eff}(\qp,L)$ is an effective two-dimensional
surface bulk modulus,
\eqn\bulktwo{B_2^{eff}(\qp,L)= B_2(\qp)+B_3(\qp)\xi_{\parallel}(\qp)
  F(L/\xi_{\parallel}),}
with $B_2(\qp)\approx B^2/(4\pi\qp)$ \refs{\mcm}.
The crossover function $F$ is given by
\eqn\crossov{F(L/\xi_{\parallel})={\beff+B_2\coth(L/\xip)
    \over B_2+\beff\coth(L/\xip)}.}
If the ratio $K/B_3$ is large, as it can indeed be at the low
fields used in the decoration experiments \ref\footc{The highest
fields probed in the decorations are of the order of $100G$.
For applied fields along the $\chat$ axis diamagnetization
corrections are large in the thin samples used in the experiments.
As a result $B\simeq H$, even for $H<H_{c1}$.},
one can distinguish three
regimes in the behavior of the effective surface bulk
modulus as a function of the in-plane wave vector $\qp$,
\eqn\bulksurf{\eqalign{B_2^{eff}(\qp,L)
    &\sim B_2(\qp)\sim\qp^{-1},~~~~~~~~\qp<{1\over L} \cr
    &\sim LB_3\sim\qp^{0},~~~~~~~~~~ {1\over L}<\qp<\qp^* \cr
    &\sim \xi_{\parallel}B_3\sim\qp^{-1},~~~~~~~~~ \qp>\qp^*. }}
The crossover wave vector $\qp^*$ is defined as the wave vector
where the correlation length $\xip$ equals the sample size,
$\xi_{\parallel}(\qp^*)=L$.
For $\qp\rightarrow 0$ the correlation length $\xip$ diverges
and the flux lines are essentially rigid.
The surface interaction dominates in this limit and
the surface bulk modulus is simply that of a 2d liquid of
magnetic monopoles. When $\qp>{1\over L}$
and $\xi_{\parallel}>L$, corresponding to $\qp<\qp^*$, flux
lines are still correlated in the $z$ direction over the entire thickness
of the sample, but
the exponential interaction between
straight flux lines in bulk dominates in this case
and the surface modulus is $LB_3$.
Finally, when $\xi_{\parallel}<L$,
or $\qp>\qp^*$, flux lines are correlated over a length $\xi_{\parallel}$
smaller than the sample size
and the surface bulk modulus is determined by the
compressional energy per
unit area of an array of essentially straight flux lines of length
$\xi_{\parallel}$.
The typical behavior of the surface structure factor as it crosses
over from a linear function of $\qp$ at very small wave vectors
to a
linear function of $\qp$ with a smaller slope at large wave vectors
is shown in Fig. 1.
For both $\qp<1/L$ and $\qp>\qp^*$
\ref\footb{We note that since we are using a hydrodynamic
theory we are always
discussing wave vectors $\qp<G_0$, where $G_0=2\pi(2/\sqrt{3}n_0)^{1/2}$
is the shortest reciprocal lattice vector of the triangular
Abrikosov lattice, corresponding
to the location of the first peak of the two-dimensional structure
function.}, the surface structure factor \stans\
is predicted to be linear in $\qp$. The dependence of the slope
on the areal density $n_0$ of flux lines is, however, different
in the two regimes. For $\qp<1/L$ we find
$S_{2T}(\qp,L)\sim \qp/n_0$, while for $\qp>\qp^*$,
$S_{2T}(\qp,L)\sim \qp/n_0^{1/2}$.
The latter behavior is consistent with
experiments.
The experiments are carried out on slabs which are
typically of $1mm$ extent and $5-30\mu m$
thickness. Due to the finite size of the decoration
images only information on the structure
function for $\qp\geq 0.4\mu m$ can typically be extracted
from the experiments that
are therefore
unable to probe the range
$\qp<1/L$. In fact the experimental structure factors appear to
level out in the small $\qp$ range \nref\murray{C.A. Murray,
private communication.}\refs{\murray}.
One can therefore conclude that the patterns probed by decorations
are indeed representative of bulk behavior.
Only the asymptotic behavior at very long length
scales is dominated by surface effects. These large length scales
have not, however, been accessible in experiments.

In Section 2 we present a simple model hydrodynamic free energy
for a flux-line liquid in a finite-size superconductor sample
that properly incorporates the boundary conditions at the
superconductor/vacuum surface. This free enrgy is used to evaluate the
thermal part of the structure function at the sample's surface.
In Section 3 we introduce disorder in the hydrodynamical
treatment and discuss its effect on the translational correlations
at the sample surface.
The comparison with experiments
is discussed in Section 4. In Appendix A we present
the results for the full three-dimensional
structure function defined in Eq. \structz .

\newsec{Hydrodynamic Free Energy and Surface Structure Function}

The pair interaction between flux lines in a
finite-size superconductor sample has been derived before in the London
approximation for both isotropic \nref\brandt{E.H. Brandt,
J. Low Temp. Phys. {\bf 42}, 557 (1981).}\refs{\brandt}
and anisotropic \nref\buisson{O. Buisson, G. Carneiro and M. Doria,
Physica C {\bf 185-189}, 1465 (1991).}\refs{\buisson ,\mcm}
superconductors.
In Ref. \refs{\mcm} we calculated the magnetic energy
of the flux-line array
in a semi-infinite sample of anisotropic $CuO_2$
superconductor occupying the half space $z<0$,
where the $z$ direction is the $\chat$ axis of the material and
the field is applied along the $\chat$ axis.
The calculation of the energy for a superconducting
slab of thickness $L$ can be carried out along the same lines.
In general the magnetic energy from the stray fields in the region ouside
the superconductor can always be rewritten as a surface contribution
to the pair interaction between flux lines that decays exponentially
with the distance from the interface.
For straight lines
the pair interaction behaves as discussed by Pearl and it decays
as $1/r_{\perp}$ at large distances at the superconductor's surfaces.
Following \refs{\mcm}, this pair interaction can then be used to obtain
the coarse-grained
hydrodynamic free energy.
The presence of the superconductor/vacuum interface modifies
the compressional and tilt elastic constants
of the flux-line array. In addition to the familiar nonlocality
of the elastic constants associated with the range of the
repulsive interaction between flux-line elements, the presence
of the interface introduces additional nonlocalities
in the $z$ direction. The corresponding surface contributions
to the elastic constants depend exponentially on the distance
of the deformed flux volume from the interface.
The surface contribution to the wave vector-dependent bulk modulus
diverges as $1/\qp$ at small wave vectors, as expected for
particles interacting via a $1/r_{\perp}$
potential in two dimensions. The surface contribution to the tilt
modulus is negative and finite in the small wavevector limit.

Instead of using this general hydrodynamic free energy,
here we propose a much simpler model free energy that
neglects all nonlocalities
in the $z$ direction, other than the presence of the
superconductor/vacuum boundaries.
For clarity we consider in this section the case of a clean
superconductor. The effect of weak impurity disorder
on the translational correlation
functions of a finite-thickness flux-line liquid
will be discussed
in the next section.
The flux liquid free energy is then written as the sum of bulk
and surface contributions,
\eqn\freet{F=F_B+F_S,}
where $F_B$ is the usual hydrodynamic free energy for a bulk flux-line
liquid. It includes terms quadratic in the hydrodynamic fields,
which are the density field
defined in \density\ and a tilt field, defined as,
\eqn\tilt{{\bf t}(\rp,z)=\sum_{i=1}^N
    {\partial\rvec_i\over\partial z}\delta(\rp-\rvec_i(z)).}
The bulk free energy is given by \refs{\mcmrev},
\eqn\freeb{F_B={1\over 2 n_0^2A}\sum_{\qpvec}\int_0^L dz\Big\{
     B_3(\qp)|\delta n(\qpvec,z)|^2
    +K(\qp) |{\bf t}(\qpvec,z)|^2  \Big\} .}
Since we are allowing for in-plane nonlocality of the elastic
constants, we have written the free energy in terms of the
Fourier components of the hydrodynamic densities.
There are two surface contributions to the free energy,
\eqn\frees{F_S={1\over 2 n_0^2A}\sum_{\qpvec}\Big\{
     B_2(\qp)|\delta n_v(\qpvec,z=0)|^2
     + B_2(\qp)|\delta n_v(\qpvec,z=L)|^2\Big\},}
where $B_2(\qp)$ is the surface contribution to the
bulk modulus obtained in \refs{\mcm} and given by
\eqn\bulks{B_2(\qp)={B^2\over 4\pi\qp}
    {1\over (1+\qp^2\lab^2)^{3/2}(\qp\lab+\sqrt{1+\qp^2\lab^2})}.}
For $\qp\lab<<1$
it is well approximated by the compressional modulus
of a two-dimensional fluid of particles interacting via a
$1/\rp$ potential,
$B_2(\qp)\simeq B^2/(4\pi\qp)$.

Statistical averages
over vortex line configurations have to be carried out
over the free energy \freet\ with the constraint that flux lines
cannot start nor stop inside the medium,
\eqn\constr{\partial_z \delta n(\qpvec,z)+
      i\qpvec\cdot{\bf t}(\qpvec,z)=0.}
It is convenient to separate ${\bf t}(\qpvec,z)$ in its
longitudinal and transverse parts,
${\bf t}=\qhat t_L+\perphat t_T$, with $t_L=\qhat\cdot{\bf t}$
and $t_T=(\perphat)\cdot{\bf t}$.
As a result of the constraint \constr , the density and the
longitudinal part of the tangent field are not independent
hydrodynamic variables. In fact they can both be expressed
expressed in terms of a ``vector potential" ${\bf u}(\qpvec,z)$,
which plays the role of a two-dimensional displacement field
for the flux-line liquid,
\eqn\vecpot{\eqalign{
  &\delta n(\qpvec,z)=-n_0i\qpvec\cdot{\bf u}(\qpvec,z) \cr
  &t_L(\qpvec,z)=n_0\partial_z\qhat\cdot{\bf u}(\qpvec,z).}}
The transverse degrees of freedom $t_T$ are decoupled from the
longitudinal ones. Since in this paper we are only interested
in calculating correlation functions of density fluctuations,
we only need to consider the longitudinal part of the free
energy. This can be expressed entirely in terms of
$u_L(\qpvec,z)=\qhat\cdot{\bf u}(\qpvec,z)$ so that
the constraint \constr\ is automatically satisfied,
with the result,
\eqn\freelong{\eqalign{F^{(L)}={1\over 2A}\sum_{\qpvec}\int_0^L&
   \bigg\{K(\qp)\Big|{du_L\over dz}\Big|^2
      + B_3(\qp)\qp^2 |u_L(\qpvec,z)|^2\cr
       &+\big[\delta(z)+\delta(z-L)\big]B_2(\qp)\qp^2|u_L(\qpvec,z)|^2
       \bigg\}.}}
The tilt energy provides the coupling between bulk and surface
terms. The model free energy of Eq. \freelong\ has the same
structure of model free energies used to study surface
phase transitions, particularly in the context of wetting
\nref\lipowsky{R. Lipowsky, Phys. Rev. Lett. {\bf 49}, 1575
(1982).}\refs{\lipowsky}.
The longitudinal
component of the vector potential takes the place of the
order parameter and the presence of interfaces introduces
spatial inhomogeneities in the order parameter.

One could now proceed directly to evaluate correlation
functions of the field $u_L$ by taking statistical averages
with the free energy \freelong . The two-dimensional structure
factor of the flux-line tips at the top surface is simply given by
$S_{2T}(\qp)=n_0\qp^2<|u_L(\qpvec,L)|^2>$, where
the angular bracket denote the statistical average over the
free energy \freelong . Since the free energy is quadratic
in the fields, these correlation functions can be
calculated exactly. On the other hand, due to the coupling of the
field at the surface to the field in the bulk of the
sample the calculation is somewhat lengthy and tedious.
A much simpler way to obtain the same result is to
use linear response theory. Let us apply a spatially
inhomogenoeus surface pressure $\delta p(\rp)$ at $z=L$ and consider
the linear response of the system to this perturbation.
The surface pressure couples to the density and the
free energy of the corresponding perturbation is given by
\eqn\press{\eqalign{\delta F_p
      & = {1\over n_0}\int d\rp
          \delta p(\rp)\delta n(\rp,L) \cr
      & = -{1\over A}\sum_{\qpvec}
       i\qp\delta p(\qpvec)u_L(\qpvec,L).}}
The response to this perturbation is a nonvanishing average
displacement, $<u_L(\qpvec,z)>_p$,
that can be determined by minimizing
the total free energy, that is, by requiring,
\eqn\minimiz{{\delta\big[F^{(L)}+\delta F_p\big]\over
    \delta u_L(\qpvec,z)} =0.}
The minimization of the total free energy yields an equation
for the displacement field in the bulk of the sample,
\eqn\mina{-K\partial^2_z<u_L(z)>_p +\qp^2 B_3<u_L(z)>_p=0,}
for $z\not=0,L$, and boundary conditions for the displacement
field at the superconductor/vacuum surfaces,
\eqn\minb{K\big[\partial_z <u_L>_p\big]_{z=L}
     +\qp^2 B_2<u_L(L)>_p -i\qp\delta p=0,}
\eqn\minc{-K\big[\partial_z <u_L>_p\big]_{z=0}
      +\qp^2 B_2<u_L(0)>_p =0.}
By solving Eq. \mina\ with the boundary conditions \minb\ and \minc\
we obtain,
\eqn\avgdis{<u_L(\qpvec,z)>_p= A_1 e^{-z/\xi_{\parallel}}
                      +A_2 e^{z/\xi_{\parallel}},}
where $\xip(\qp)$ is the correlation length defined in Eq. \xipar\
and
\eqn\const{\eqalign{&A_1={i\over 2\qp}{B_3\xi_{\parallel}-B_2\over
     [B_3^2\xi^2_{\parallel}+B^2_2]\sinh(L/\xi_{\parallel})
   +2B_3\xi_{\parallel}B_2\cosh(L/\xi_{\parallel})}~\delta p(\qpvec),\cr
                    &A_2={i\over 2\qp}{B_3\xi_{\parallel}+B_2\over
     [B_3^2\xi^2_{\parallel}+B^2_2]\sinh(L/\xi_{\parallel})
   +2B_3\xi_{\parallel}B_2\cosh(L/\xi_{\parallel})}~\delta p(\qpvec).}}
The surface displacement is then given by,
\eqn\surfdis{<u_L(\qpvec,L)>_p=i{\delta p(\qpvec)\over\qp}~{1\over
      B_2(\qp)+B_3(\qp)\xip(\qp)F(L/\xip)},}
where $F(L/\xip)$ is the crossover function given in Eq. \crossov .
As expected, the response to the applied pressure
is linear
in the perturbation. The corresponding
linear response function defined as
\eqn\compress{\chi_T(\qp,L)=
     {\delta n(\qpvec,L)\over\delta p(\qpvec)}
     = {-i\qp n_0<u_L(\qpvec,L)>_p\over\delta p(\qpvec)},}
is the wave vector-dependent surface
isothermal compressibility, which is in turn the reciprocal
of the two-dimensional surface
bulk modulus, $B_2^{eff}(\qp,L)=1/\chi(\qp,L)$.
The denominator of Eq. \surfdis\ is then the effective surface
bulk modulus given in Eq. \bulktwo .
Finally, the surface structure factor is given by Eq. \stans .

\newsec{Weak Disorder}

The effect of weak disorder both in the bulk of the sample
and at the superconductor/vacuum interfaces can be modeled
in a standard way in terms of a random potential that couples
to the density field. In general surface disorder may
include surface roughness and therefore differ considerably
from impurity disorder in bulk.
We therefore model bulk and surface disorder separately
by adding to the free energy given in Eq. \freelong\ two terms,
\eqn\freedis{\delta F_D=\int d\rp\int_0^L
   \Big\{V_b(\rp,z)\delta n(\rp,z)
    +[\delta(z)+\delta(z-L)]V_s(\rp)\delta n(\rp,z)\Big\},}
where the random potentials $V_b(\rp,z)$ and $V_s(\rp)$
represent the effect of random impurities and small scale
inhomogeneities in the bulk and at the surface, respectively.
If the defects are randomly distributed, as for instance in the case of
oxygen vacancies, we expect that the quenched fluctuations in the
impurity potentials will obey,
\eqn\imppot{\eqalign{& \overline{V_b(\rp,z)V_b(\rp',z')}=\Delta_b
          \delta(\rp-\rp')\delta(z-z'), \cr
      & \overline{V_s(\rp)V_s(\rp')}=\Delta_s
          \delta(\rp-\rp') ,}}
with $\overline{V_b(\rp,z)}=0$ and $\overline{V_s(\rp)}=0$.
%
%

It was shown by Nelson and Le Doussal \refs{\doussal}
that in bulk flux-line liquids a hydrodynamic treatment of
weak bulk disorder produces
``Lorentzian squared" correction to the hydrodynamic result
for the structure factor, given by
\eqn\structdis{S_{bd}(\qp,q_z)=n_0\Delta_b\bigg(
    {n_0\qp^2\over \qp^2B_3+q_z^2K}\bigg)^2,}
or, by inverting the $q_z$ transform,
\eqn\strucbz{S_{bd}(\qp,z_1-z_2)=n_0\Delta_b\Big({n_0\over
     B_3(\qp)\xip(\qp)}
   \Big)^2\big[\xip(\qp)+|z_1-z_2|\big]e^{-|z_1-z_2|/\xip}.}
The contribution from weak disorder to two-dimensional density
correlations in a constant-$z$ cross-section of bulk,
is then obtained from \strucbz\ by letting $z_1=z_2$,
\eqn\structwob{S_{2bd}(\qp)=
       {n_0^3\Delta_b\xip(\qp)\over [B_3(\qp)\xip(\qp)]^2}.}
Notice that this contribution to the two dimensional
structure factor of a cross section of flux-line liquid also
vanishes linearly with $\qp$ at small wave vectors.

To evaluate the two-dimensional structure factor of flux-line
tips at the surface of a slab of type-II superconductor  in a
field, we proceed as in Section 2. The total free energy,
$F^{(L)}+F_D$,
of the flux-line liquid consists now of the sum of Eqs. \freelong\
and \freedis .
To this we add the surface perturbation given in \press\
and evaluate the response of the system by minimizing the total
free energy.
As in Section 2 we obtain an
equation for the displacement field in the bulk of the sample,
\eqn\minad{-K\partial^2_z<u_L(z)>_p
    +\qp^2 B_3<u_L(z)>_p-i\qp n_0V_b(\qpvec,z)=0,}
for $z\not=0,L$, and boundary conditions for the displacement
field at the superconductor/vacuum surfaces,
\eqn\minbd{K\big[\partial_z <u_L>_p\big]_{z=L}
     +\qp^2 B_2<u_L(L)>_p -i\qp n_0V_s(\qpvec)-i\qp\delta p=0,}
\eqn\mincd{-K\big[\partial_z <u_L>_p\big]_{z=0}
      +\qp^2 B_2<u_L(z=0)>_p -i\qp n_0V_s(\qpvec)=0.}
The solution of Eq. \minad\ with the boundary conditions
\minbd\ and \mincd\ is given by,
\eqn\disdis{<u_L(\qpvec,z)>_p=A'_1e^{-z/\xip}+A'_2e^{z/\xip}
         -{in_0\xip\over K}\int_0^z dz'\cosh[(z-z')/\xip]
         V_b(\qpvec,z'),}
with
\eqn\coeffdisa{\eqalign{A'_1={i\over 2\qp}&{1\over
     [B_3^2\xi^2_{\parallel}+B^2_2]\sinh(L/\xi_{\parallel})
   +2B_3\xi_{\parallel}B_2\cosh(L/\xi_{\parallel})}
     \bigg\{[B_3\xi_{\parallel}-B_2]\delta p(\qpvec)\cr
    & +n_0V_s(\qpvec)\big[(B_3\xi_{\parallel}-B_2)
       +(B_3\xi_{\parallel}+B_2)e^{L/\xip}\big] \cr
    & +n_0(B_3\xi_{\parallel}-B_2)\int_0^L dzV_b(\qpvec,z)
     \Big[\cosh\Big({L-z\over\xip}\Big)
     +{B_2\over B_3\xip}\sinh\Big({L-z\over\xip}\Big) \Big]
      \bigg\},}}
\eqn\coeffdisb{\eqalign{A'_2={i\over 2\qp}&{1\over
     [B_3^2\xi^2_{\parallel}+B^2_2]\sinh(L/\xi_{\parallel})
   +2B_3\xi_{\parallel}B_2\cosh(L/\xi_{\parallel})}
     \bigg\{[B_3\xi_{\parallel}+B_2]\delta p(\qpvec)\cr
    & +n_0V_s(\qpvec)\big[(B_3\xi_{\parallel}+B_2)
       +(B_3\xi_{\parallel}-B_2)e^{-L/\xip}\big] \cr
    & +n_0(B_3\xi_{\parallel}+B_2)\int_0^L dzV_b(\qpvec,z)
     \Big[\cosh\Big({L-z\over\xip}\Big)
     +{B_2\over B_3\xip}\sinh\Big({L-z\over\xip}\Big) \Big]
      \bigg\}.}}
The solution \disdis\ depends explicitly on the impurity potentials
$V_b$ and $V_s$ and represents the longitudinal
displacement field for a given realization of
the disorder. In the presence of quenched disorder the
fluctuation-dissipation theorem no longer holds and one
cannot simply identify the ratio of the mean displacement
$\overline{<u_L(\qpvec,z)>_p}$ to the perturbation
$\delta p(\qpvec)$ with the
system's compressibility. In fact the
mean displacement in the presence of weak disorder is simply
equal to that given in \avgdis\ for a clean superconductor.
This is because the terms linear
in the impurity potentials in Eq. \disdis\
vanish when averaged over the quenched disorder
and $\overline{A'_1}=A_1$ and $\overline{A'_2}=A_2$.
One then needs to calculate directly the correlation
function of the fluctuations as defined in Eq. \threestr\ ,
or, in terms
of the linear response, $<u_L(\qpvec,L)>_p$,
\eqn\structd{S(\qp,z_1,z_2)=n_0\qp^2\Big[
    \overline{<u_L(\qpvec,z_1)>_p<u_L(-\qpvec,z_2)>}
    - \overline{<u_L(\qpvec,z_1)>_p}~~\overline{<u_L(-\qpvec,z_2)>_p}
     \Big].}
After some algebra, one finds that the two-dimensional
structure factor at the surface $z=L$ of the slab,
$S_2(\qp,L)=S(\qp,z_1=L,z_2=L)$, can be written as
the sum of three contributions,
\eqn\surfst{S_2(\qp,L)=S_{2T}(\qp,L)+S_{2db}(\qp,L)+S_{2sd}(\qp,L),}
where $S_{2T}(\qp,L)$ is the contribution from thermal fluctuations
that was already obtained in Section 2,
\eqn\surfstrt{S_{2T}(\qp,L)={n_0k_BT\over B_2^{eff}(\qp,L)},}
with $B_2^{eff}(\qp,L)$ the surface bulk modulus given in Eq.
\bulktwo .
The other two terms in Eq. \surfst\ represent the contributions
from bulk and surface disorder, respectively. They are given by,
\eqn\surfstrbd{\eqalign{S_{2bd}(\qp,L)= &
     {n_0^3\Delta_b\xip\over 2\big[B_2^{eff}(\qp,L)\big]^2}
     {1\over \big[B_2+\beff\coth(L/\xip)\big]^2} \cr
 &  \times \bigg\{\big[B_3^2\xip^2+B_2^2\big]\coth(L/\xip)
    +2B_2B_3\xip+\big[B_3^2\xip^2-B_2^2\big]{L/\xip\over\sinh^2(L/\xip)}
        \bigg\},}}
and
\eqn\surfstrsd{S_{2sd}(\qp,L)=
     {n_0^3\Delta_s\over \big[B_2^{eff}(\qp,L)\big]^2}
     \bigg\{1+{B_3^2\xip^2/\sinh^2(L/\xip)\over
        \big[B_2+B_3\xip\coth(L/\xip)\big]^2}\bigg\}.}
While these expressions appear rather complicated, one can show
that for all $\qp$ of interest here the terms inside curly brackets
in Eqs. \surfstrbd\ and \surfstrsd\ are of order one.

\newsec{Discussion}

The surface structure factors extracted from the analysis
of the low-field decoration experiments carried out by the
AT\&T group show clearly a large peak at
$\qp=4\pi/(\sqrt{3}a_0)$, where $a_0=(2/\sqrt{3}a_0)^{1/2}$
is the nearest neighbor vortex spacing in the triangular
Abrikosov lattice.
corresponding to the nearest-neighbor vortex spacing \refs{\gammel}.
At smaller wave vectors the structure functions obtained at different
applied fields all decrease linearly with
wave vector according to $S_2(\qp,L)\sim\qp/n_0^{1/2}$ \refs{\gammel}.
At the smallest wave vectors probed in the experiments,
the decay of the structure functions seem to be levelling
out \refs{\murray}.
This behavior is most apparent in the experiments at the
lowest fields.

To discuss the comparison of our results
with these experimental findings,
it is useful to consider
some limiting forms of the expressions
obtained in sections 2 and 3.

When the sample thickness $L$ is large compared to the correlation
length $\xip$ the expressions given in sections 2 and 3 for
the surface structure factors reduce to those for the two-dimensional
structure factors at the surface of a semi-infinite sample
of type-II superconductor.
In this limit, $L>>\xip$ or $\qp>>\qp^*$, the crossover function given
in Eq. \crossov\ is $F(L/\xip)\approx 1$ and we find
\eqn\stapprox{S_{2T}(\qp,L)\simeq {n_0k_BT\over
           B_2+\beff}\sim \qp,}
\eqn\sbdapprox{S_{2bd}(\qp,L)\simeq {n_0^3\Delta_b\xip\over
          2\big[ B_2+\beff]^2}\sim \qp,}
\eqn\ssdapprox{S_{2sd}(\qp,L)\simeq {n_0^3\Delta_s\over
          \big[ B_2+\beff]^2}\sim \qp^2.}
In this limit all contributions to the surface structure factor
are determined by an effective bulk modulus
$B_2+\beff\sim B_3(1+\sqrt{K/B_3})/\qp$, where we approximated
$B_2\sim B_3/\qp$ \refs{\mcm}.
As indicated in Eqs. \stapprox - \ssdapprox ,
while both the contributions to the structure factor from thermal
fluctuations and from bulk disorder are linear in $\qp$ in this
regime, the contribution from surface disorder is
proportional to $\qp^2$.
We argue below that the linear
dependence of the surface structure factor on $\qp$ observed
in the decoration experiments by the AT\&T group \refs{\gammel ,\murray}
probably corresponds to this regime.
If our interpretation is correct, our results indicate
that surface roughness does not play an important role in
these experiments.

In the opposite limit, $L<<\xip$, the flux lines are essentially
straight over the entire thickness of the sample.
The crossover function \crossov\ becomes
$F(L/\xip)\approx(B_2+B_3L)/\beff$ and the
various contributions to the surface structure factor
are given by,
\eqn\stappr{S_{2T}(\qp,L)\simeq {n_0k_BT\over
           2B_2+B_3L},}
\eqn\sbdappr{S_{2bd}(\qp,L)\simeq {n_0^3\Delta_bL\over
          2\big[2 B_2+B_3L]^2},}
\eqn\ssdappr{S_{2sd}(\qp,L)\simeq {2n_0^3\Delta_s\over
          \big[2 B_2+B_3L]^2}.}
When $\qp\rightarrow 0$, $B_2\sim\qp^{-1}$ while
$B_3\sim\qp^0$ and
$B_2$ always dominates in the denominator of Eqs.
\stappr -\ssdappr , yielding,
\eqn\stzero{\lim_{\qp\rightarrow 0}S_{2T}(\qp,L)
          \simeq{n_0k_BT\over 2B_2}\sim\qp,}
\eqn\sbdzero{\lim_{\qp\rightarrow 0}S_{2bd}(\qp,L)
    \simeq{n_0^3\Delta_bL\over 4B_2^2}\sim\qp^2,}
\eqn\ssbdzero{\lim_{\qp\rightarrow 0}S_{2sbd}(\qp,L)
     \simeq{n_0^3\Delta_s\over 2B_2^2}\sim\qp^2.}
If $B_2\sim B_3/\qp$, the above limiting forms apply in the entire
range $\qp<<1/L$.
Finally, if the ratio $K/B_3$ is large enough so that there is
a nonvanishing range of wavevectors where $1/L<\qp<\qp^*$,
in this range the surface structure factors level out to a
constant value, given by,
\eqn\stconst{S_{2T}(\qp,L)\simeq {n_0k_BT\over LB_3},}
\eqn\sbdconst{S_{2bd}(\qp,L)\simeq {n_0^3\Delta_b\over LB_3^2},}
\eqn\ssdconst{S_{2sd}(\qp,L)\simeq {2n_0^3\Delta_s\over (LB_3)^2}.}
We remark that if one evaluates the surface structure factors for a
finite-thickness superconducting slab using free boundary conditions
for the flux lines at the sample surfaces, one finds
that in the limit
$\qp\rightarrow 0$ the various contributions to $S_2(\qp,L)$
have precisely the sample-size-dependent constant values given
in \stconst -\ssdconst .
In fact for all values of $\qp$ where our
hydrodynamic theory is relevant,
the expressions for $S_2(\qp,L)$ for the case of free boundary
conditions on the flux lines are obtained from those given here
by setting $B_2=0$.
The precise nature of the boundary conditions
only affects the very small $\qp$ behavior of the correlation functions.

We conclude by discussing in some detail the comparison of our
findings with the experimental structure factors obtained
by C.A. Murray and collaborators from low field decoration
images \refs{\gammel ,\murray}.
The fields in these experiments are in the range $8-50G$,
corresponding to flux-line areal densities of $0.4-2.5\mu m^{-2}$.
The smallest wave vectors at which one can construct a structure
function from the decoration images is $\sim 0.4\mu m^{-1}$.
The long wavelength tail of $S_2(\qp)$ is found to fit a $\qp/n^{1/2}_0$
in an intermediate range of  wave vectors. The slope of $S_2(\qp)$
in this linear region is typically
$\sim 5\times 10^{-3}-10^{-2}\mu m$. At the smallest wave vectors where
data are available $S_2(\qp)$ appears to be levelling out to a
constant value $\sim 0.02$, almost independent of the
applied field.

Our model predicts a linear decrease of $S_{2T}(\qp)$ with $\qp$
in the limit $\qp\rightarrow 0$. The corresponding slope is
determined by $B_2\sim n_0^2\phi_0^2/(4\pi\qp)$
according to Eq. \stzero .
One finds $S_{2T}(\qp)\sim 2\times 10^{-6}\mu m^{-1}K^{-1}(T/n_0)\qp$.
Using $T\sim 88K$ and $n_0\sim 1\mu m^{-2}$, we obtain
$S_{2T}(\qp)\sim 2\times 10^{-5}\mu m~\qp$, a slope over two order
of magnitude smaller than oberved in experiments
In addition the dependence of the slope on the density is not
of the form obtained in the experimental fit.
Finally, in the limit $\qp\rightarrow 0$ both contributions
to the structure functions from weak disorder vanish
as $\qp^2$. All of this is consistent with the
expectation that this asymptotic long wavelength regime
is never probed in the experiments.

If the ratio $\sqrt{K/B_3}\leq 1$, then for all wave vectors accessible
to experiments $\xip<L$ and the various contributions to the
surface structure function are given by Eqs. \stapprox -\ssdapprox .
Again we have a linear dependence of $S_{2T}(\qp)$ on $\qp$,
but with a slope even smaller than in the asymptotic
$\qp\rightarrow 0$ case discussed above.
A similar estimate can be carried out for the contribution
from bulk disorder. In this case the size of the linear
slope is determined by $\Delta_b$. If the pinning is by isolated
$O_2$ vacancies, we can estimate $\Delta_b\sim U_0^2\xi_{ab}$,
with $U_0\sim 5K$ a typical pinning energy barrier.
The corresponding linear slope is then even smaller than for the
thermal part of the correlation function.

Our results are qualitatively consistent with the experimental
findings if $\sqrt{K/B_3}>> 1$. In this case we predict a
crossover from the linear dependence of $S_2(\qp)$ on $\qp$
described by Eqs. \stapprox\ and \sbdapprox\ for $\qp>\qp^*$
to the sample-size dependent constant values given by Eqs. \stconst\
and \sbdconst . On the other hand, the predicted size of
the constant value
of $S_2(\qp)$ for $\qp\leq\qp^*$ and the linear slope at
larger wave vectors are more than an order of magnitude
smaller than from experiments, even when we take into account
the in-plane nonlocality of the elastic constants.
We can in turn try to fit our results to experiments to extract
experimental values for the elastic constants.
For instance in Fig. 1 by requiring
$S_2(\qp\geq 1/L)\sim n_0k_BT/(B_3^{exp}L)\sim 0.02$ for $H=8G$,
we find $B_3^{exp}/k_B\sim 70K\mu m^{-3}$ for $L\sim 10\mu m$.
This value is about $500$ times smaller than
$B_3(0)=B^2/4\pi$, nor can the in-plane nonlocality of $B_3$
account for the difference.
We can then extract a value of the tilt modulus $K$ by
fitting the slope in the linear region,
$S_2(\qp\geq\qp^*)\sim(n_0k_BT/\sqrt{K^{exp}B_3^{exp}})\qp\sim
0.02\mu m~\qp$. We find $K^{exp}/k_B\sim 4.4\times
10^4 k\mu m^{-3}$, a value not inconsistent with the
theoretical value for $K(0)$. The corresponding crossover
wave vector is $\qp^*=(1/L)\sqrt{K^{exp}/B_3^{exp}}\sim 1.\mu m^{-1}$.
A mechanism that can account quantitatively for the very small value
of $B_3$ could be the strong downward renormalization of
the compressional modulus that was predicted by Nelson and
Le Doussal at low density \refs{\doussal}.

In conclusion, our work indicates clearly that, while the decay of
translational correlations at very long wavelengths
($\qp\rightarrow 0$) is indeed governed by surface
properties and finite size effects, the correlation
functions extracted from the decoration experiments
are in fact representative of bulk behavior.
More data at small wave vectors and low fields with
a detailed analysis of such data are, however,
needed for a quantitative comparison.

\bigskip
Work by M.C.M. was supported by the NSF through grant DMR-91-12330.
D.R.N. acknowledges support from the NSF through the Harvard Materials
Research Laboratory and through grant DMR-91-15491.
We are grateful to C.A. Murray and D. Grier for sharing with
me experimental results prior to publication.

\appendix{A}{Three-dimensional Structure Function}

In this Appendix we discuss of the full three-dimensional
structure function given in Eqs. \structz .
We only sketch the derivation of the thermal part of this
correlation function, $S_T(\qp,z,z_0)$.
This derivation is instructive as an
example of how to evaluate correlation functions directly
by taking statistical averages with the free energy \freelong\
rather than by using linear response theory.

It is convenient to expand the displacement field
$u_L(\qpvec,z)$ in a Fourier series,
\eqn\disfou{u_L(\qpvec,z)={1\over L}\Big\{u_0(\qpvec)
   +2\sum_{p=1}^{\infty}\big[u_p(\qpvec)\cos(p\pi z/L)
       +v_p(\qpvec)\sin(p\pi z/L)\big]\Big\}.}
The free energy \freelong\ is immediately rewritten in
terms of the Fourier amplitudes of the displacement
field, with the result,
\eqn\freefou{\eqalign{F^{L}={1\over 2AL^2}\sum_{\qpvec}
  \bigg\{&\sum_{p=-\infty}^{\infty}\alpha_p(\qp)
       \big(|u_p(\qpvec)|^2+|v_p(\qpvec)|^2\big)\cr
   &  +{B_2\over L}\qp^2\bigg|\sum_{p=-\infty}^{\infty}
     u_p(\qpvec)\bigg|^2
     +{B_2\over L}\qp^2\bigg|\sum_{p=-\infty}^{\infty}
        (-1)^p u_p(\qpvec)\bigg|^2\bigg\},}}
with
\eqn\coeffdef{\alpha_p(\qp)=K\big({p\pi\over L}\big)^2
   +B_3\qp^2,}
where $v_0(\qpvec)=0$ and we used that
$u_p(-\qpvec)=u_p(\qpvec)$ and $v_p(-\qpvec)=-v_p(\qpvec)$.
Using Eq. \vecpot ,
the thermal part of the three-dimensional structure function
can be written in terms of the longitudinal
displacement field,
\eqn\structdd{S_T(\qp,z_1,z_2)=n_0\qp^2
    <u_L(\qpvec,z_1)u_L(-\qpvec,z_2)>.}
By inserting Eq. \disfou\ into Eq. \structdd\ we obtain
\eqn\structfou{\eqalign{S_T(\qp,z_1,z_2)={n_0\qp^2\over L^2}
       \sum_{p=-\infty}^{\infty} \sum_{p'=-\infty}^{\infty} &
    \Big\{<u_p(\qp)u_{p'}^*(\qp)>\cos(p\pi z_1/L)
       \cos(p'\pi z_2/L)\cr
  & +  <v_p(\qp)v_{p'}^*(\qp)>\sin(p\pi z_1/L)
       \sin(p'\pi z_2/L)\Big\}.}}
The correlation functions of the the Fourier amplitudes
of the displacement field can be calculated by using standard
tricks to deal with coupled gaussian integrals. The result is
\eqn\oddamp{<v_p(\qpvec)v_{p'}(-\qpvec)>=
         \delta_{p,p'}{Lk_BT\over\alpha_p(\qp)},}
and
\eqn\evamp{\eqalign{<u_p(\qpvec)u_{p'}(-\qpvec)>= &
         \delta_{p,p'}{Lk_BT\over\alpha_p(\qp)}
     -{k_BT B_2\qp^2\over\alpha_p\alpha_{p'}}
     {1\over (1+B_2\qp^2S_o)^2-(B_2\qp^2S_e)^2}\cr
    & \times\Big\{(1+B_2\qp^2S_o)[1+(-1)^{p+p'}]
     -B_2\qp^2S_e[(-1)^p+(-1)^{p'}]\Big\}.}}
The quantities denoted by $S_o$ and $S_e$ are special cases
of more general sums that will be needed below. These are,
\eqn\sumev{\Sigma_o(\qp,z)={1\over L}\sum_{p=-\infty}^{\infty}
             {\cos(p\pi z/L)\over\alpha_p(\qp)}
        ={1\over\beff \qp^2}{\cosh[(z-L)/\xip]\over
            \sinh(L/\xip)},}
\eqn\sumodd{\Sigma_e(\qp,z)={1\over L}\sum_{p=-\infty}^{\infty}
             {(-1)^p\cos(p\pi z/L)\over\alpha_p(\qp)}
           ={1\over\beff \qp^2}{\cosh(z/\xip)\over
            \sinh(L/\xip)},}
with $\xip(\qp)$ the correlation length given in Eq. \xipar .
The quantities $S_o$ and $S_e$ are given by the values of the
above sums at $z=0$, i.e., $S_o=\Sigma_o(\qp,0)$ and
$S_e=\Sigma_e(\qp,0)$.

Finally, by inserting Eqs. \evamp\ and \oddamp\ into Eq. \structfou ,
one obtains,
\eqn\structtot{\eqalign{S_T(\qp,&z_1,z_2)={n_0k_BT\over\beff}
   \bigg\{ {\cosh[(z_1-z_2-L)/\xip]\over\sinh(L/\xip)}
      -{B_2\over B_2+\beff\xip F(L/\xip)}\cr
      &\times \bigg[ {\cosh[(z_1+z_2-L)/\xip]\over\sinh(L/\xip)}
    +{\beff\over B_2+\beff\coth(L/\xip)}~
     {\cosh[(z_1-z_2)/\xip]\over\sinh^2(L/\xip)}
    \bigg]\bigg\} .}}
The two-dimensional structure factor in a constant-$z$ cross-section
is obtained from Eq. \structtot\ by letting $z_1=z_2$,
\eqn\structtwoz{\eqalign{S_{2T}(\qp,z_1)={n_0k_BT\over\beff} &
      {1\over B_2+\beff\xip F(L/\xip)}\bigg\{\beff+B_2\coth(L/\xip)\cr
    &  -{B_2\beff\over B_2+\beff\coth(L/\xip)}~
       {\cosh[(2z_1-L)/\xip]\over\sinh(L/\xip)}\bigg\}.}}
For $z_1=L$ and $z_1=0$ this is identical to the espression given
in Eq. \stans .

It is interesting to consider the correlation of density
fluctuations on the two opposite surfaces of the sample,
corresponding to
the three dimensional structure function of \structtot\ for
$z_1=L$ and $z_2=0$. From Eq. \structtot\ we obtain
\eqn\stracr{S_T(\qp,L,0)=S_{2T}(\qp,0)R(\qp,L),}
where $S_{2T}(\qp,0)=S_{2T}(\qp,L)$ is
the two-dimensional structure factor of one of the two surfaces,
and
\eqn\ratios{R(\qp,L)={\beff\over B_2\sinh(L/\xip)+\beff\cosh(L/\xip)}}
measures the correlations between flux-line patterns
at the two opposite ends of the sample.
If the size of
the sample is small compared to the correlation length, $L<<\xip$,
flux lines are straight throughout the sample and $R(\qp,L)\approx 1$.
The patterns
on the two surfaces are perfectly correlated in this limit.
Conversely, if $L>>\xip$ the patterns
on the two surface are uncorrelated and $R(\qp,L)\approx 0$.
A measure of the deviation of $R(\qp,L)$ from $1$ would give us
information about the degree of flux-line wandering in the
superconductor.

\vfill\eject
\listrefs

\vfil\eject
\figures

\fig{1}{The thermal contribution to the surface structure function, as
given in Eq. \stb , for $H=8G$ (for this value of the field
the first maximum of $S_2(\qp)$
is at $\qp\simeq 4.26\mu m^{-1}$), $T=88K$, $\lab=0.3\mu m$, $\gamma=55$
and $L=25\mu m$.
Here $B_3(0)$ was treated as a parameter determined by fitting
our results to the
small wave vector part of the data of C.A. Murray et al. \refs{\murray},
as discussed in Section 3.}

\end